\documentstyle[12pt]{article}
\begin{document}
\begin{flushright}
WU-AP/41/95
\end{flushright}
\vskip 0.5cm
\baselineskip = 18pt
\begin{center}
{\large{\bf BUBBLES WITH AN {\it\bf O}(3) SYMMETRIC SCALAR FIELD\\
IN CURVED SPACETIME}}\footnote{
Talk given at 7th Marcel Grossmann Meeting on General Relativity,
Stanford University, July 24-29, 1994.}

\vskip 0.5cm
{\sc Nobuyuki Sakai$^{(1)}$, Yoonbai KIM$^{(2)}$, and Kei-ichi
MAEDA$^{(1)}$}\\  $^{(1)}$ {\it Department of Physics, Waseda
University, Shinjuku-ku,  Tokyo 169, Japan}\\  $^{(2)}$ {\it Department
of Physics, Pusan National University, Pusan 609-735, Korea} \end{center}

\vskip 0.3cm
\begin{abstract}

\baselineskip = 18pt
We study the first-order phase transition in a model of scalar field
with $O(3)$ symmetry coupled to gravity, and, in high temperature limit,
discuss the existence of new bubble solution with a global monopole at
the center of the bubble.
\end{abstract}

\vskip 0.5cm
Since the first-order phase transition was formulated in the context of
Euclidean path integral \cite{1}, it has attracted much attention as a
possible resolution of cosmological problems. It is widely believed that
the first-order phase transition is described by the formation and
growth of bubbles in which no matter lumps remain.

In this note, we shall discuss a possibility of a new
bubble solution that the global monopole is
supported at the center of it \cite{2,3}. Such a new
type of solution was first discovered in a flat
spacetime by one of the present authors \cite{2}. The
$O(3)$-symmetric scalar field in the presence of
gravity at high temperature is described by the action,
\vskip -0.5cm $~$
$$
S_E=\int^{{1\over T}}_{0}\! dt\int\!
d^3x\!\sqrt{g}\,\Bigl\{-{R\over16\pi G}
+{1\over2}g^{\mu\nu}
\partial_{\mu}\phi^{a}\partial_{\nu}\phi^{a}+V(\phi)
\Bigr\}.
$$
In our calculation we assume a sixth-order scalar potential such as
$$
V(\phi)=\lambda\Bigl({\phi^2 \over
v^2} +\alpha\Bigr)(\phi^{2}-v^{2})^{2},
{}~~~~(0<\alpha<0.5)
$$
however, our argument  does scarcely depend on the detailed form
of  scalar potential if it has a true vacuum at $\phi_{-}$ with
$V(\phi_{-})=0$ and a false vacuum at $\phi_{+}$ with
$V(\phi_{+})>0$. Since   time-dependence can be
neglected in high-temperature limit, the
spherically symmetric bubble solutions are constituted
under the hedgehog ansatz,
$\phi^{a}=\phi(r)(\sin n\theta\cos n\varphi,\,\sin
n\theta\sin n\varphi, \,\cos n\theta) $, where $n$ has
to be 0 or 1 for the sake of the spherical symmetry.
We present a numerical solution of scalar field
in Fig.1 and distribution of the energy
density in Fig.2 for $\lambda=1,~ \alpha=0.1,~
v=0.1 M_{Pl}$. Fig.1 shows that the $n=0$ bubble  is
nothing but the normal high-temperature bubble in a
real singlet scalar model \cite{4} where its central
region is in true vacuum, but the center of the $n=1$
bubble should be in false vacuum because of the
winding between  $O(3)$ internal symmetry and spatial
rotation.  From the profiles of energy density $\rho$
in Fig.2, we find that a  global monopole is formed
at the center of the $n=1$ bubble. This global monopole
is surrounded by the inner wall
$(R/H^{-1}\sim 0.1)$ which distinguishes the lump
from the true vacuum region $(0.2<R/H^{-1}<0.5)$. The
long-range tail of global monopole, $ \rho \sim
1/r^{2}$, renders the spacetime between the inner
and outer walls flat with the deficit (solid)
angle $\Delta=8\pi^{2}Gv^{2}$. When we consider the
object in Planck scale, the spacetime of monopole
will be closed by itself.
However, if we consider a non-Abelian gauge field as
well, just as in the standard model or grand unified
theories, we will find a formation of black hole
carrying a magnetic charge \cite{5}.

We now turn to the evaluation of nucleation rate,
$\Gamma\sim Ae^{-B}$.
Since the $n=1$ bubble consumes additional energy to
support a global monopole, its size   is larger
than that of $n=0$ bubble (see Fig.2), and then
$B_{1}$ is always larger than $B_{0}$
regardless of the strength of gravity and of
the shape of the scalar potential, where $B_n$
is the value of Euclidean action of $n$ bubble.
As an example, we show a numerical result of
$B'\equiv(T/v)B$ in Fig.3 ($\lambda=1,~\alpha=0.1$) and Fig.4
($\lambda=1,~v=0.1M_{pl}$). The exponential part of  nucleation rate,
$\Gamma^{(n)}\sim e^{-B_{n}}$, tells us that the $n=0$
bubble is always more favorable than the $n=1$ bubble.
However, we find that gravity enhances
the decay rate, which is consistent with the results
for a singlet scalar field at zero temperature \cite{6}, and for large
$\alpha$ (small potential barrier), the decay rate of
the $n=1$ bubble becomes comparable to that of
the $n=0$ bubble, i.e., the relative decay rate
$\Gamma^{(1)}/\Gamma^{(0)}$ may be determined by the ratio of prefactors
$A_n$ (see Ref.2).  We may need further analysis including high
temperature quantum correction effect on the potential  to give a
definite answer.

Though the high-temperature bubbles are given by
static solutions when they are nucleated, they will
immediately start to grow by the following reasons;
one is some combustion process when the environment
keeps the temperature  high \cite{7,4} and the other is
the recovery of zero-temperature classical dynamics due to the expansion
of background universe dominated by radiation. We only take into account
the latter effect here. The motion of outer wall of $n=1$ bubble
resembles that of $n=0$ bubble. For the matter droplet inside $n=1$
bubble, the global monopole is stable under the influence of gravitation
when the phase transition scale is lower than Planck scale, as shown in
Figs.5 and 6 ($\lambda=1,~\alpha=0.2,~v=0.1$). In Fig.6 we present
trajectories of positions of $\phi=0.5v$. However, if we consider this
object in Planck scale, an issue that whether the core site shows a
topological inflation \cite{8} or it shrinks to a black hole or a
wormhole \cite{9} needs further study.

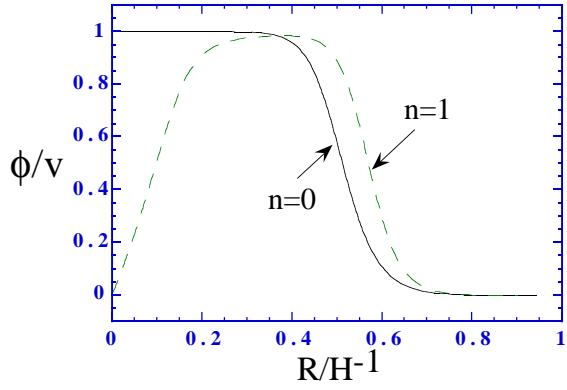
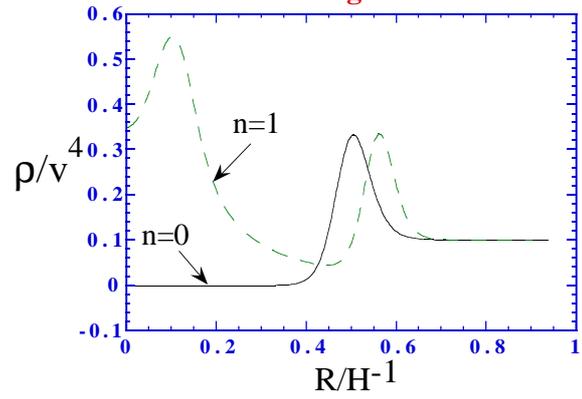
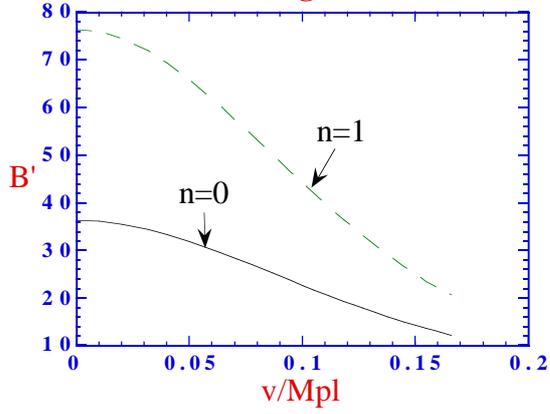
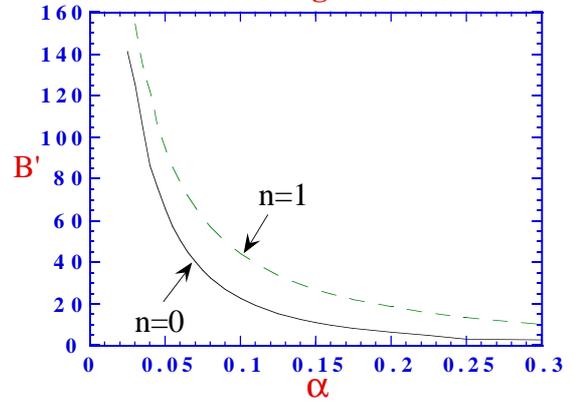
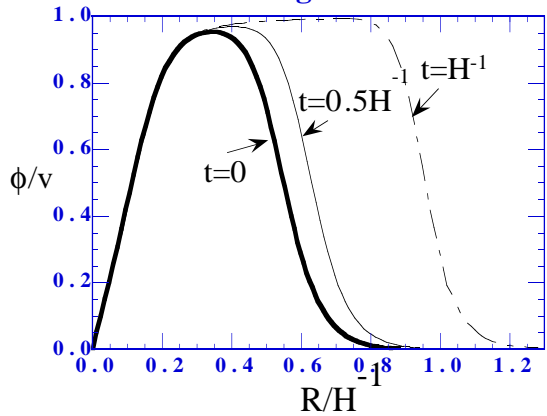
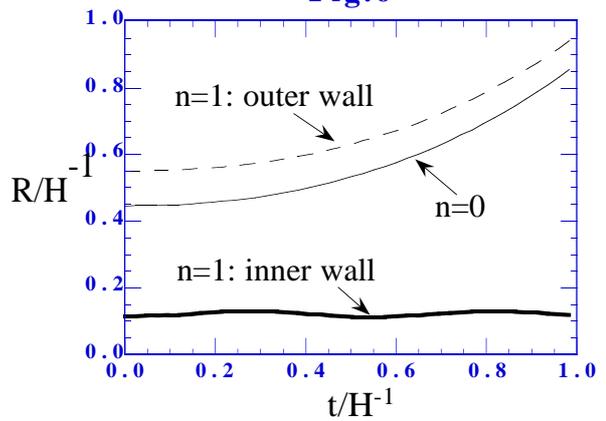
\end{document}